\documentclass{article}
\usepackage{amsmath,amssymb,epsfig}

\begin{document}

\thispagestyle{empty} \vspace*{1cm} \rightline{Napoli DSF-T-39/2004} %
\rightline{INFN-NA-39/2004} \vspace*{2cm}

\begin{center}
{\LARGE Topological order in Josephson junction ladders with
Mobius boundary conditions}

{\LARGE \ }

\vspace{8mm}

{\large Gerardo Cristofano\footnote{{\large {\footnotesize
Dipartimento di Scienze Fisiche,}{\it \ {\footnotesize
Universit\'{a} di Napoli ``Federico II''\ \newline and INFN,
Sezione di Napoli}-}{\small Via Cintia - Compl.\ universitario M.
Sant'Angelo - 80126 Napoli, Italy}}}, Vincenzo
Marotta\footnotemark[1]  , }

{\large Adele Naddeo\footnote{{\large {\footnotesize Dipartimento
di Scienze Fisiche,}{\it \ {\footnotesize Universit\'{a} di Napoli
``Federico II''
\newline
and INFM, Unit\`{a} di Napoli}-}{\small Via Cintia - Compl.
universitario M. Sant'Angelo - 80126 Napoli, Italy}}} }

{\small \ }

{\bf Abstract\\[0pt]
}
\end{center}

\begin{quotation}
We propose a CFT description for a closed one-dimensional fully
frustrated
ladder of quantum Josephson junctions with Mobius boundary conditions \cite%
{noi}, in particular we show how such a system can develop
topological order. Such a property is crucial for its
implementation as a ``protected'' solid state qubit.

\vspace*{0.5cm}



{\footnotesize Work supported in part by the European Communities
Human Potential}

{\footnotesize Program under contract HPRN-CT-2000-00131 Quantum
Spacetime\newpage } \setcounter{page}{2}
\end{quotation}

\section{Introduction}

The concept of topological order was first introduced to describe the ground
state of a quantum Hall fluid \cite{wen}. Although todays interest in
topological order mainly derives from the quest for exotic non-Fermi liquid
states relevant for high $T_{c}$ superconductors \cite{fisher}, such a
concept is of much more general interest \cite{wen1}.

Two features of topological order are very striking: fractionally charged
quasiparticles and a ground state degeneracy depending on the topology of
the underlying manifold, which is lifted by quasiparticles tunnelling
processes. For Laughlin fractional quantum Hall (FQH) states both these
properties are well understood \cite{wen2}, but for superconducting devices
the situation is less clear.

Josephson junctions networks appear to be good candidates for
exhibiting topological order, as recently evidenced in Refs.
\cite{ioffe}\cite{pasquale} by means of Chern-Simons gauge field
theory. Such a property may allow for their use as ``protected''
qubits for quantum computation. In this paper we shall show that
fully frustrated Josephson junction ladders (JJL) with non trivial
geometry may support topological order, making use of conformal
field theory techniques \cite{noi}. A simple experimental test of
our predictions will be also proposed.

The paper is organized as follows.

In Section 2 we introduce the fully frustrated quantum Josephson
junctions ladder (JJL) focusing on non trivial boundary
conditions.

In Section 3 we recall some aspects of the $m$-reduction procedure \cite{VM}%
, in particular we show how the $m=2$, $p=0$ case of our twisted
model (TM) \cite{cgm2} well accounts for the symmetries of the
model under study. In such a framework we give the whole primary
fields content of the theory on the plane and exhibit the ground
state wave function.

In Section 4, starting from our CFT results, we show that the ground state
is degenerate, the different states being accessible by adiabatic flux
change techniques. Such a degeneracy is shown to be strictly related to the
presence in the spectrum of quasiparticles with non abelian statistics and
can be lifted non perturbatively through vortices tunnelling.

In Section 5 some comments and outlooks are given.

In the Appendix we recall briefly the boundary states introduced in Ref.
\cite{noi1} in the framework of our TM.

\section{Josephson junctions ladders with Mobius boundary conditions}

In this Section we briefly describe the system we will study in the
following, that is a closed ladder of Josephson junctions (see Fig.1) with
Mobius boundary conditions. With each site $i$ we associate a phase $\varphi
_{i}$ and a charge $2en_{i}$, representing a superconducting grain coupled
to its neighbours by Josephson couplings; $n_{i}$ and $\varphi _{i}$ are
conjugate variables satisfying the usual phase-number commutation relation.
The system is described by the quantum phase model (QPM) Hamiltonian:
\begin{equation}
H=-\frac{E_{C}}{2}\sum_{i}\left( \frac{\partial }{\partial \varphi _{i}}%
\right) ^{2}-\sum_{\left\langle ij\right\rangle }E_{ij}\cos \left( \varphi
_{i}-\varphi _{j}-A_{ij}\right) ,  \label{act0}
\end{equation}
where $E_{C}=\frac{\left( 2e\right) ^{2}}{C}$ ($C$ being the capacitance) is
the charging energy at site $i$, while the second term is the Josephson
coupling energy between sites $i$ and $j$ and the sum is over nearest
neighbours. $A_{ij}=\frac{2\pi }{\Phi _{0}}$ $\int_{i}^{j}A{\cdot }dl$ is
the line integral of the vector potential associated to an external magnetic
field $B$ and $\Phi _{0}=\frac{hc}{2e}$ is the superconducting flux quantum.
The gauge invariant sum around a plaquette is $\sum_{p}A_{ij}=2\pi f$ with $%
f=\frac{\Phi }{\Phi _{0}}$, where $\Phi $ is the flux threading each
plaquette of the ladder.

\begin{figure}[tbp]
\centering\includegraphics*[width=0.6\linewidth]{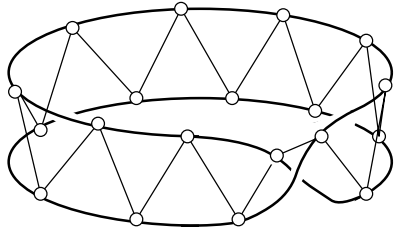}
\caption{Josephson junction ladder with Mobius boundary conditions}
\label{figura1}
\end{figure}

Let us label the phase fields on the two legs with $\varphi _{i}^{\left(
a\right) }$, $a=1,2$ and assume $E_{ij}=E_{x}$ for horizontal links and $%
E_{ij}=E_{y}$ for vertical ones. Let us also make the gauge choice $%
A_{ij}=+\pi f$ for the upper links, $A_{ij}=-\pi f$ for the lower ones and $%
A_{ij}=0$ for the vertical ones, which corresponds to a vector potential
parallel to the ladder and taking opposite values on upper and lower
branches.

Thus the effective quantum Hamiltonian (\ref{act0}) can be written
as \cite{ladder}:
\begin{eqnarray}
-H &=&\frac{E_{C}}{2}\sum_{i}\left[ \left( \frac{\partial }{\partial \varphi
_{i}^{\left( 1\right) }}\right) ^{2}+\left( \frac{\partial }{\partial
\varphi _{i}^{\left( 2\right) }}\right) ^{2}\right] +  \nonumber \\
&&\sum_{i}\left[ E_{x}\sum_{a=1,2}\cos \left( \varphi _{i+1}^{\left(
a\right) }-\varphi _{i}^{\left( a\right) }+\left( -1\right) ^{a}\pi f\right)
+E_{y}\cos \left( \varphi _{i}^{\left( 1\right) }-\varphi _{i}^{\left(
2\right) }\right) \right] .  \label{ha1}
\end{eqnarray}

The correspondence between the effective quantum Hamiltonian (\ref{ha1}) and
our TM model can be best traced performing the change of variables \cite%
{ladder}: $\varphi _{i}^{\left( 1\right) }=X_{i}+\phi _{i}$, $\varphi
_{i}^{\left( 2\right) }=X_{i}-\phi _{i}$, so getting:
\begin{eqnarray}
H &=&-\frac{E_{C}}{2}\sum_{i}\left[ \left( \frac{\partial }{\partial X_{i}}%
\right) ^{2}+\left( \frac{\partial }{\partial \phi _{i}}\right) ^{2}\right] -
\nonumber \\
&&\sum_{i}\left[ 2E_{x}\cos \left( X_{i+1}-X_{i}\right) \cos \left( \phi
_{i+1}-\phi _{i}-\pi f\right) +E_{y}\cos \left( 2\phi _{i}\right) \right]
\label{ha2}
\end{eqnarray}
where $X_{i}$, $\phi _{i}$ (i.e. $\varphi _{i}^{\left( 1\right) }$, $\varphi
_{i}^{\left( 2\right) }$) are only phase deviations of each order parameter
from the commensurate phase and should not be identified with the phases of
the superconducting grains \cite{ladder}.

When $f=\frac{1}{2}$ and $E_{C}=0$ (classical limit) the ground
state of the $1D$ frustrated quantum XY (FQXY) model displays - in
addition to the continuous $U(1)$ symmetry of the phase variables
- a discrete $Z_{2}$ symmetry associated with an antiferromagnetic
pattern of plaquette chiralities $\chi _{p}=\pm 1$, measuring the
two opposite directions of the supercurrent circulating in each
plaquette. The evidence for a chiral phase in Josephson junction
ladders has been investigated in Ref. \cite{nishi} while a field
theoretical description of chiral order is developed in
\cite{azaria}.

Performing the continuum limit of the Hamiltonian (\ref{ha2}):
\begin{eqnarray}
-H &=&\frac{E_{C}}{2}\int dx\left[ \left( \frac{\partial }{\partial X}%
\right) ^{2}+\left( \frac{\partial }{\partial \phi }\right) ^{2}\right] +
\nonumber \\
&&\int dx\left[ E_{x}\left( \frac{\partial X}{\partial x}\right)
^{2}+E_{x}\left( \frac{\partial \phi }{\partial x}-\frac{\pi }{2}\right)
^{2}+E_{y}\cos \left( 2\phi \right) \right]  \label{ha3}
\end{eqnarray}
we see that the $X$ and $\phi $ fields are decoupled. In fact the $X$ term
of the above Hamiltonian is that of a free quantum field theory while the $%
\phi $ one coincides with the quantum sine-Gordon model. Through an
imaginary-time path-integral formulation of such a model it can be shown
that the $1D$ quantum problem maps into a $2D$ classical statistical
mechanics system, the $2D$ fully frustrated XY model, where the parameter $%
\alpha =\left( \frac{E_{x}}{E_{C}}\right) ^{\frac{1}{2}}$ plays
the role of an inverse temperature \cite{ladder}. For small
$E_{C}$ there is a gap for creation of kinks in the
antiferromagnetic pattern of $\chi _{p}$ and the ground state has
quasi long range chiral order. We work in the regime
${E_{x}}\gg{E_{y}}$ where the ladder is well described by a CFT
with central charge $c=2$.

We are now ready to introduce the modified ladder \cite{noi}, see
Fig. 1. In order to do so let us first require the $\varphi
^{\left( a\right) }$, $a=1,2$, variables to recover the angular
nature by compactification of both the up and down fields. In such
a way the XY-vortices, causing the Kosterlitz-Thouless transition,
are recovered. As a second step let us introduce at a point $x=0$
a defect which couples the up and down edges through its
interaction with the two legs, that is let us close the ladder and
impose Mobius boundary conditions. In the limit of strong coupling
such an interaction gives rise to non trivial boundary conditions
for the fields \cite{noi1}.
 In the following we give further details on such
an issue, in particular we adopt the $m$-reduction technique \cite{VM}\cite%
{cgm2}, which accounts for non trivial boundary conditions
\cite{noi1} for the Josephson ladder in the presence of a defect
line. In the Appendix the relevant chiral fields ${{\varphi}_e}
^{\left( a\right) }$, $a=1,2$, which emerge from such conditions,
are explicitly constructed, by using the folding procedure.

\section{$m$-reduction technique}

In this Section we focus on the $m$-reduction technique for the
special $m=2$ case and apply it to the system described by the
Hamiltonian (\ref{ha3}). In the Appendix each phase field $\varphi
^{\left( a\right) }$ is written as a sum of two fields of opposite
chirality defined on an half-line, because of the presence of a
defect at $x=0$. Within a ''bosonization'' framework it is shown
there how it is possible to reduce to a problem with two chiral
fields $\varphi _{e}^{\left( a\right) }$, $a=1,2$, each defined on
the whole $x-$axis, and the corresponding dual fields. Now we
identify in the continuum such chiral phase fields $\varphi
_{e}^{\left( a\right) }$, $a=1,2$, each defined on the
corresponding leg, with the two chiral fields $Q^{\left( a\right)
}$, $a=1,2$ of our CFT, the TM, with central charge $c=2$.

In order to construct such fields we start from a CFT with $c=1$ described
in terms of a scalar chiral field $Q$ compactified on a circle with radius $%
R^{2}=2$, explicitly given by:
\begin{equation}
Q(z)=q-i\,p\,lnz+i\sum_{n\neq 0}\frac{a_{n}}{n}z^{-n}  \label{modes}
\end{equation}
with $a_{n}$, $q$ and $p$ satisfying the commutation relations $\left[
a_{n},a_{n^{\prime }}\right] =n\delta _{n,n^{\prime }}$ and $\left[ q,p%
\right] =i$; its primary fields are the vertex operators $U^{\alpha }\left(
z\right) =:e^{i\alpha Q(z)}:$. It is possible to give a plasma description
through the relation $\left| \psi \right| ^{2}=e^{-\beta H_{eff}}$ where $%
\psi \left( z_{1},...,z_{N}\right) =\left\langle N\alpha
|\prod_{i=1}^{N}U^{\alpha }(z_{i})|0\right\rangle =\prod_{i<j=1}^{N}\left(
z_{i}-z_{j}\right) ^{2}$ is the ground state wave function. It can be
immediately seen that $H_{eff}=-\sum_{i<j=1}^{N}\ln \left|
z_{i}-z_{j}\right| $ and $\beta =\frac{2}{R^{2}}=1$, that is only vorticity $%
v=1$ vortices are present in the plasma.

Starting from such a CFT mother theory one can use the $m$-reduction
procedure, which consists in considering the subalgebra generated only by
the modes in eq. (\ref{modes}) which are a multiple of an integer $m$, so
getting a $c=m$ orbifold CFT (daughter theory, i.e. the twisted model (TM))
\cite{cgm2}. With respect to the special $m=2$ case, the fields in the
mother CFT can be organized into components which have well defined
transformation properties under the discrete $Z_{2}$ (twist) group, which is
a symmetry of the TM. By using the mapping $z\rightarrow z^{1/2}$ and by
making the identifications $a_{2n+l}\longrightarrow \sqrt{2}a_{n+l/2}$, $%
q\longrightarrow \frac{1}{\sqrt{2}}q$ the $c=2$ daughter CFT is obtained. It
is interesting to notice that such a daughter CFT gives rise to a vortices
plasma of half integer vorticity, that is to a fully frustrated XY model, as
it will appear in the following.

Its primary fields content can be expressed in terms of a $Z_{2}$-invariant
scalar field $X(z)$, given by
\begin{equation}
X(z)=\frac{1}{2}\left( Q^{(1)}(z)+Q^{(2)}(z)\right) ,  \label{X}
\end{equation}
describing the continuous phase sector of the new theory, and a twisted
field
\begin{equation}
\phi (z)=\frac{1}{2}\left( Q^{(1)}(z)-Q^{(2)}(z)\right) ,  \label{phi}
\end{equation}
which satisfies the twisted boundary conditions $\phi (e^{i\pi }z)=-\phi (z)$
\cite{cgm2}. Such fields coincide with the ones introduced in eq. (\ref{ha3}%
).

The whole TM theory decomposes into a tensor product of two CFTs,
a twisted invariant one with $c=\frac{3}{2}$ and the remaining
$c=\frac{1}{2}$ one realized by a Majorana fermion in the twisted
sector. In the $c=\frac{3}{2}$ subtheory the primary fields are
composite vertex operators $V\left( z\right) =U_{X}\left( z\right)
\psi \left( z\right) $ or $V_{qh}\left( z\right) =U_{X}\left(
z\right) \sigma \left( z\right) $, where $U_{X}\left( z\right)
=\frac{1}{\sqrt{z}}:e^{i\alpha X(z)}:$ is the vertex of the
charged\ sector with $\alpha ^{2}=2$ for the $SU(2)$ Cooper
pairing symmetry used here.

Regarding the other\ component, the highest weight state in the neutral
sector can be classified by the two chiral operators:
\begin{eqnarray}
\psi \left( z\right) &=&\frac{1}{2\sqrt{z}}\left( :e^{i\alpha \phi \left(
z\right) }:+:e^{i\alpha \phi \left( -z\right) }:\right) ,~~~~~~  \nonumber \\
\overline{\psi }\left( z\right) &=&\frac{1}{2\sqrt{z}}\left( :e^{i\alpha
\phi \left( z\right) }:-:e^{i\alpha \phi \left( -z\right) }:\right) ;
\label{neu11}
\end{eqnarray}
which correspond to two $c=\frac{1}{2}$ Majorana fermions with Ramond
(invariant under the $Z_{2}$ twist) or Neveu-Schwartz ($Z_{2}$ twisted)
boundary conditions \cite{cgm2} in a fermionized version of the theory. Let
us point out that the energy-momentum tensor of the Ramond part of the
neutral sector develops a cosine term:
\begin{equation}
T_{\psi }\left( z\right) =-\frac{1}{4}\left( \partial \phi \right) ^{2}-%
\frac{1}{16z^{2}}\cos \left( 2\sqrt{2}\phi \right) ,  \label{tn1}
\end{equation}
a clear signature of a tunneling phenomenon which selects a new stable
vacuum, the linear superposition of the two ground states. The Ramond fields
are the degrees of freedom which survive after the tunneling and the $Z_{2}$
(orbifold) symmetry, which exchanges the two Ising fermions, is broken.

So the whole energy-momentum tensor within the $c=\frac{3}{2}$ subtheory is:
\begin{equation}
T=T_{X}\left( z\right) +T_{\psi }\left( z\right) =-\frac{1}{2}\left(
\partial X\right) ^{2}-\frac{1}{4}\left( \partial \phi \right) ^{2}-\frac{1}{%
16z^{2}}\cos \left( 2\sqrt{2}\phi \right) .  \label{ttt1}
\end{equation}
The correspondence with the Hamiltonian in eq. (\ref{ha3}) is more evident
once we observe that the neutral current $\partial \phi $ appearing above
coincides with the term $\partial \phi -\frac{\pi }{2}$ of eq. (\ref{ha3}),
since the $\frac{\pi }{2}$-term coming there from the frustration condition,
here it appears in $\partial \phi $ as a zero mode, i.e. a classical mode.
Besides the fields appearing in eq. (\ref{neu11}) there are the $\sigma
\left( z\right) $ fields, also called the twist fields, which appear in the
primary fields $V_{qh}\left( z\right) $ combined to a vertex with charge $%
\frac{e}{4}$. The twist fields have non local properties and decide also for
the non trivial properties of the vacuum state, which in fact can be twisted
or not in our formalism. Such a property for the vacuum is more evident for
the torus topology, where the $\sigma $-field is described by the conformal
block $\chi _{\frac{1}{16}}$ (see Section 4).

The evidence of a phase transition in ladder systems at
$c={\frac{3}{2}}$ has been investigated in \cite{gritsev} within a
CFT framework. Within this framework the ground state wavefunction
is described as a correlator of $N_{2e}$ Cooper pairs:
\begin{equation}
<N_{2e}\alpha |\prod_{i=1}^{N_{2e}}V^{\sqrt{2}}(z_{i})|0>=\prod_{i<i^{\prime
}=1}^{N_{2e}}(z_{i}-z_{i^{\prime }})Pf\left( \frac{1}{z_{i}-z_{i^{\prime }}}%
\right)  \label{pff}
\end{equation}%
where $Pf\left( \frac{1}{z_{i}-z_{i^{\prime }}}\right) ={\cal A}\left( \frac{%
1}{z_{1}-z_{2}}\frac{1}{z_{3}-z_{4}}\dots \right) $ is the antisymmetrized
product over pairs of Cooper pairs, so reproducing well known results \cite%
{MR}. In a similar way we also are able to evaluate correlators of $N_{2e}$
Cooper pairs in the presence of (quasi-hole) excitation \cite{MR}\cite{cgm2}
with non Abelian statistics \cite{nayak}. It is now interesting to notice
that the charged contribution appearing in the correlator of $N_{e}$
electrons is just: $<N_{e}\alpha
|\prod_{i=1}^{N_{2e}}U_{X}^{1/2}(z_{i})|0>=\prod_{i<i^{\prime
}=1}^{N_{e}}(z_{i}-z_{i^{\prime }})^{1/4}$, giving rise to a vortices plasma
with $H_{eff}=-\frac{1}{4}\sum_{i<j=1}^{N}\ln \left\vert
z_{i}-z_{j}\right\vert $ at the corresponding "temperature" $\beta =2$, that
is it describes vortices with vorticity $v=\frac{1}{2}$!

On closed annulus geometries, as it is the discretized analogue of a torus,
we must properly account for boundary conditions at the ends of the finite
lattice since they determine in the continuum the pertinent conformal blocks
yielding the statistics of quasiparticles as well as the ground state
degeneracy. We have two possible boundary conditions which correspond to two
different ways to close the double lattice, i.e. $\varphi _{N}^{\left(
a\right) }\rightarrow \varphi _{1}^{\left( a\right) }$ or $\varphi
_{N}^{\left( a+1\right) }\rightarrow \varphi _{1}^{\left( a\right) }$, $%
a=1,2 $. It is not difficult to work out that, for the ladder case, the
twisted boundary conditions can be implemented only on the odd lattice.

\begin{figure}[tbp]
\centering\includegraphics*[width=0.9\linewidth]{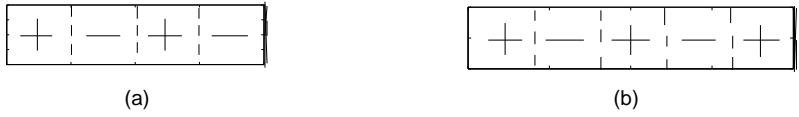}
\caption{boundary conditions: (a) untwisted; (b) twisted. }
\label{figura2}
\end{figure}

Indeed for $f=\frac{1}{2}$ the ladder is invariant under the shift of two
sites, so that there are two topologically inequivalent boundary conditions
for even or odd number of sites. In the even case the end sites are of the
same kind of the starting one, while in the odd case a ferromagnetic line
corresponds to an antiferromagnetic one. The even (odd) case corresponds to
untwisted (twisted) boundary conditions, as depicted in Fig. 2. The odd case
selects out two degenerate ground states which are in different topological
sectors, so the system may develop topological order in the twisted sector
of our theory. Let us notice that in the discrete case not all the vacua in
the different sectors of our theory are connected. For instance the states
in the untwisted sector, which correspond to the even ladder, are physically
disconnected from those of the twisted one, which correspond to the odd
ladder.

\section{Topological order and ``protected'' qubits}

The aim of this Section is to fully exploit the issue of topological order
in a quantum JJL. In order to meet such a request let us use the results of
the $2$-reduction technique for the torus topology \cite{cgm2}. On the torus
the TM primary fields are described in terms of the conformal blocks of the $%
c=\frac{3}{2}$ subtheory and an Ising model, so reflecting the decomposition
on the plane outlined in the previous Section. The following characters,
expressed in terms of the torus variable $w=\frac{1}{2\pi i}\ln z$,
\begin{eqnarray}
\chi _{(0)}^{c=3/2}(w|\tau ) &=&\chi _{0}(\tau )K_{0}\left( w|\tau \right)
+\chi _{\frac{1}{2}}(\tau )K_{2}\left( w|\tau \right)  \label{mr1} \\
\chi _{(1)}^{c=3/2}(w|\tau ) &=&\chi _{\frac{1}{16}}(\tau )\left(
K_{1}\left( w|\tau \right) +K_{3}\left( w|\tau \right) \right)  \label{mr2}
\\
\chi _{(2)}^{c=3/2}(w|\tau ) &=&\chi _{\frac{1}{2}}(\tau )K_{0}\left( w|\tau
\right) +\chi _{0}(\tau )K_{2}\left( w|\tau \right)  \label{mr3}
\end{eqnarray}%
represent the field content of the $Z_{2}$ invariant $c=3/2$ \ CFT with a
\textquotedblleft charged\textquotedblright\ component ($K_{\alpha }(w|\tau
) $, see definition given below) and a \textquotedblleft
neutral\textquotedblright\ component ($\chi _{\beta }$, the conformal blocks
of the Ising Model). In order to understand the physical significance of the
$c=2$ conformal blocks in terms of the charged low energy excitations of the
system, let us evidence their electric charge (and magnetic flux contents in
the dual theory, obtained by exchanging the compactification radius $%
R_{e}^{2}\rightarrow R_{m}^{2}$ in the charged sector of the CFT). Hence let
us consider the \textquotedblleft charged\textquotedblright\ sector
conformal blocks appearing in eqs. (\ref{mr1}-\ref{mr3}):
\begin{equation}
K_{2l+i}(w|\tau )=\frac{1}{\eta \left( \tau \right) }\Theta \left[
\begin{array}{c}
\frac{2l+i}{4} \\
0%
\end{array}%
\right] (2w|4\tau ),~~~~~\forall \left( l,i\right) \in \left( 0,1\right)
^{2},
\end{equation}%
corresponding to primary fields with conformal dimensions $h_{2l+i}=\frac{1}{%
2}\alpha _{\left( l,i\right) }^{2}=\frac{1}{2}\left( \frac{2l+i}{2}+2\delta
_{\left( l+i\right) ,0}\right) ^{2}$ and electric charges $2e\left( \frac{%
\alpha _{\left( l,i\right) }}{R_{X}}\right) $ (magnetic charges in the dual
theory $\frac{hc}{2e}\left( \alpha _{\left( l,i\right) }R_{X}\right) $), $%
R_{X}=1$ being the compactification radius.

Now we turn to the whole $c=2$ theory. The characters of the twisted sector
are given by:
\begin{eqnarray}
\chi _{(0)}(w|\tau ) &=&\bar{\chi}_{\frac{1}{16}}\left( \chi
_{(0)}^{c=3/2}(w|\tau )+\chi _{(2)}^{c=3/2}(w|\tau )\right)  \label{tw1} \\
&=&\bar{\chi}_{\frac{1}{16}}\left( \chi _{0}+\chi _{\frac{1}{2}}\right)
\left( K_{0}+K_{2}\right)  \nonumber \\
\chi _{(1)}(w|\tau ) &=&\left( \bar{\chi}_{0}+\bar{\chi}_{\frac{1}{2}%
}\right) \chi _{(1)}^{c=3/2}(w|\tau )  \label{tw2} \\
&=&\left( \bar{\chi}_{0}+\bar{\chi}_{\frac{1}{2}}\right) \chi _{\frac{1}{16}%
}\left( K_{1}+K_{3}\right)  \nonumber
\end{eqnarray}
where $\bar{\chi}_{\beta }$ are the $c=\frac{1}{2}$ Ising characters. Such a
factorization is a consequence of the parity selection rule ($m$-ality),
which gives a gluing condition for the ``charged''\ and ``neutral''\
excitations. Furthermore the characters of the untwisted sector are \cite%
{cgm2}:
\begin{eqnarray}
\tilde{\chi}_{(0)}^{+}(w|\tau ) &=&\bar{\chi}_{0}\chi _{(0)}^{c=3/2}(w|\tau
)+\bar{\chi}_{\frac{1}{2}}\chi _{(2)}^{c=3/2}(w|\tau )  \label{vac1} \\
&=&\left( \bar{\chi}_{0}\chi _{0}+\bar{\chi}_{\frac{1}{2}}\chi _{\frac{1}{2}%
}\right) K_{0}+\left( \bar{\chi}_{0}\chi _{\frac{1}{2}}+\bar{\chi}_{\frac{1}{%
2}}\chi _{0}\right) K_{2}  \nonumber \\
\tilde{\chi}_{(1)}^{+}(w|\tau ) &=&\bar{\chi}_{0}\chi _{(2)}^{c=3/2}(w|\tau
)+\bar{\chi}_{\frac{1}{2}}\chi _{(0)}^{c=3/2}(w|\tau )  \label{vac2} \\
&=&\left( \bar{\chi}_{0}\chi _{\frac{1}{2}}+\bar{\chi}_{\frac{1}{2}}\chi
_{0}\right) K_{0}+\left( \bar{\chi}_{0}\chi _{0}+\bar{\chi}_{\frac{1}{2}%
}\chi _{\frac{1}{2}}\right) K_{2}  \nonumber \\
\tilde{\chi}_{(0)}^{-}(w|\tau ) &=&\bar{\chi}_{0}\chi _{(0)}^{c=3/2}(w|\tau
)-\bar{\chi}_{\frac{1}{2}}\chi _{(2)}^{c=3/2}(w|\tau )  \label{vac3} \\
&=&\left( \bar{\chi}_{0}\chi _{0}-\bar{\chi}_{\frac{1}{2}}\chi _{\frac{1}{2}%
}\right) K_{0}+\left( \bar{\chi}_{0}\chi _{\frac{1}{2}}-\bar{\chi}_{\frac{1}{%
2}}\chi _{0}\right) K_{2}  \nonumber \\
\tilde{\chi}_{(1)}^{-}(w|\tau ) &=&\bar{\chi}_{0}\chi _{(2)}^{c=3/2}(w|\tau
)-\bar{\chi}_{\frac{1}{2}}\chi _{(0)}^{c=3/2}(w|\tau )  \label{vac4} \\
&=&\left( \bar{\chi}_{0}\chi _{\frac{1}{2}}-\bar{\chi}_{\frac{1}{2}}\chi
_{0}\right) K_{0}+\left( \bar{\chi}_{0}\chi _{0}-\bar{\chi}_{\frac{1}{2}%
}\chi _{\frac{1}{2}}\right) K_{2}  \nonumber \\
\tilde{\chi}_{(0)}(w|\tau ) &=&\bar{\chi}_{\frac{1}{16}}\chi
_{(1)}^{c=3/2}(w|\tau )=\bar{\chi}_{\frac{1}{16}}\chi _{\frac{1}{16}}\left(
K_{1}+K_{3}\right) .  \label{vac5}
\end{eqnarray}

The conformal blocks given above represent the collective states of highly
correlated vortices, which appear to be incompressible.

For closed geometries the JJL with Mobius boundary conditions
gives rise to a line defect in the bulk. So it becomes mandatory
to use a folding procedure to map the problem with a defect line
into a boundary one, where the defect line appears as a boundary
state. The TM boundary states have been constructed in Ref.
\cite{noi1} together with the corresponding chiral partition
functions and briefly recalled in the Appendix. In particular we
get an \textquotedblright untwisted\textquotedblright\ sector and
a \textquotedblright twisted\textquotedblright\ one, corresponding
to periodic and Mobius boundary conditions respectively. That
gives rise to an essential difference in the low energy spectrum
for the system under study:
in the first case only two-spinon excitations are possible (i.e. ${SU(2)}%
_{2} $ integer spin representations) while, in the last case, the presence
of a topological defect provides a clear evidence of single-spinon
excitations (i.e. ${SU(2)}_{2}$ half-integer spin representations). All that
takes place in close analogy with spin-1/2 closed zigzag ladders, which are
expected to map to fully frustrated Josephson ladders in the extremely
quantum limit. In this way all previous results obtained by means of exact
diagonalization techniques \cite{mobius} are recovered.

In the following we discuss topological order referring to the characters
which in turn are related to the different boundary states present in the
system through such chiral partition functions \cite{noi}. We can build a
topological invariant ${\cal P}=\prod\limits_{\gamma }\chi _{p}$ where $%
\gamma $ is a closed contour that goes around the hole. Such a choice allows
us to define two degenerate ground states with ${\cal P}=1$ and ${\cal P}=-1$
respectively, labelled $\left\vert 1\right\rangle ,\left\vert 2\right\rangle
$, as in Fig. 3(a): there must be an odd number of plaquettes (odd ladder)
to satisfy this condition which is imposed by the request of topological
protection. The value ${\cal P}=-1$ selects twisted boundary conditions at
the ends of the chain while ${\cal P}=1$ selects periodic boundary
conditions (see Fig. 3(b)). The case of finite discrete systems has been
discussed in detail in Ref. \cite{grimm}, our theory is its counterpart in
the continuum.

\begin{figure}[tbp]
\centering\includegraphics*[width=0.7\linewidth]{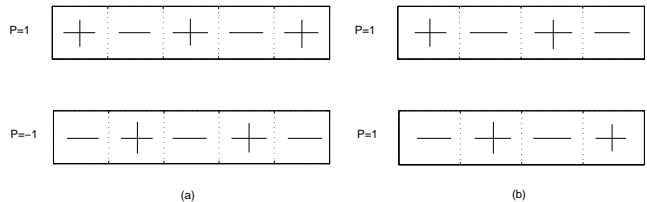}
\caption{degenerate ground states for: (a) odd ladder, ${\cal P}=\pm 1$; (b)
even ladder, ${\cal P}=1$. }
\label{figura3}
\end{figure}

It is now possible to identify the two degenerate ground states shown in
Fig.3(a) with the characters (\ref{p6}), (\ref{p7}) in the twisted sector of
our theory (see Appendix) and then to remove such a degeneracy through
vortices tunneling: the last operation is also needed in order to prepare
the qubit in a definite state.

We recognize such two degenerate ground states as the ones
corresponding to the twisted characters (\ref{tw1}-\ref{tw2}):
\begin{equation}
\left\vert 0\right\rangle \rightarrow \chi _{(0)}(w|\tau ),\left\vert
1\right\rangle \rightarrow \chi _{(1)}(w|\tau ).  \label{gsd}
\end{equation}%
Upon performing an adiabatic change of local magnetic fields which drags one
half vortex across the system, i.e. through the transport of a half flux
quantum around the $B$-cycle of the torus, we are able to flip the state of
the system, so lifting the degeneracy.

Such two states can be distinguished through the presence or absence
respectively of a half flux quantum trapped inside the central hole which,
in turn, is related to the presence or absence of the Ising character $\chi
_{\frac{1}{16}}$ in the $c=\frac{3}{2}$ subtheory. The trapped half flux
quantum can be experimentally detected, so giving a way to read out the
state of the system. Furthermore it should be noticed that the presence of
the twist operators $\sigma $, described here by the Ising character $\chi _{%
\frac{1}{16}}$, gives rise to non Abelian statistics, which can be evidenced
by their fusion rules $\sigma \sigma ={\bf 1}+\psi $ \cite{MR}\cite{cgm2}.

\section{Conclusions and outlooks}

In this paper we have shown that Josephson junctions ladders with
non trivial geometry may develop topological order allowing for
the implementation of \textquotedblleft
protected\textquotedblright\ qubits, a first step toward the
realization of an ideal solid state quantum computer. Josephson
junctions
ladders with annular geometry have been fabricated within the trilayer $%
Nb/Al-AlO_{x}/Nb$ technology and experimentally investigated \cite{ustinov}.
So in principle it could be simple to conceive an experimental setup in
order to test our predictions.

\section{Appendix}

In such Appendix we recall briefly the TM boundary states (BS)
recently constructed in \cite{noi1}. For closed geometries, that
is for the torus, the JJL with an impurity gives rise to a line
defect in the bulk. In a theory with a defect line the interaction
with the impurity gives rise to the following non trivial boundary
conditions for the fields:
\begin{equation}
\varphi _{L}^{\left( a\right) }\left( x=0\right) =\mp \varphi
_{R}^{\left( a\right) }\left( x=0\right) -\varphi _{0},\text{ \ \
}a=1,2.  \label{blr}
\end{equation}%
In order to describe it we resort to the
folding procedure. Such a procedure is used in the literature to
map a problem with a defect line (as a bulk property) into a
boundary one, where the defect line appears as a boundary state of
a theory which is not anymore chiral and its fields are defined in
a reduced region which is one half of the original one. Our
approach, the TM, is a chiral description of that, where the
chiral $\phi $\ field defined in ($-L/2$, $L/2)$ describes both
the left moving component and the right moving one defined in ($-L/2$, $\ 0$%
), ($0$, $L/2$) respectively, in the folded description
\cite{noi1}. Furthermore to make a connection with the TM we
consider more general gluing conditions:

$\phi _{L}(x=0)=\mp \phi _{R}(x=0)-\varphi _{0}$

the $-$($+$) sign staying for the twisted (untwisted) sector. We
are then allowed to use the boundary states given in
\cite{Affleck} for the $c=1$ orbifold at the Ising$^{2}$ radius.
The $X$ field, which is even under the
folding procedure, does not suffer any change in boundary conditions \cite%
{noi1}. Let us now write each phase field as the sum $\varphi
^{\left( a\right) }\left( x\right) =\varphi _{L}^{\left( a\right)
}\left( x\right) +\varphi _{R}^{\left( a\right) }\left( x\right) $
of left and right moving fields defined on the half-line because
of the defect located in $x=0$. Then let us define for each leg
the two chiral fields $\varphi _{e,o}^{\left( a\right) }\left(
x\right) =\varphi _{L}^{\left( a\right) }\left( x\right) \pm
\varphi _{R}^{\left( a\right) }\left( -x\right) $, each defined on
the whole $x-$axis \cite{boso}. In such a framework the dual
fields $\varphi _{o}^{\left( a\right) }\left( x\right) $ are fully
decoupled because the corresponding boundary interaction term in
the Hamiltonian does not involve them \cite{affleck}; they are
involved in the definition of the conjugate momenta $\Pi _{\left(
a\right) }=\left( \partial _{x}\varphi _{o}^{\left( a\right)
}\right) =\left( \frac{\partial }{\partial \varphi _{e}^{\left(
a\right) }}\right) $ present in the quantum Hamiltonian.
Performing the change of variables $\varphi _{e}^{\left( 1\right)
}=X+\phi $, $\varphi
_{e}^{\left( 2\right) }=X-\phi $ ($\varphi _{o}^{\left( 1\right) }=\overline{%
X}+\overline{\phi }$, $\varphi _{o}^{\left( 2\right) }=\overline{X}-%
\overline{\phi }$ for the dual ones) we get the quantum Hamiltonian (\ref%
{ha3}) but, now, all the fields are chiral ones.

It is interesting to notice that the condition (\ref{blr}) is
naturally satisfied by the twisted field $\phi \left( z\right) $
of our twisted model (TM) (see eq. (\ref{phi})).

The most convenient representation of such BS is the one in which
they appear as a product of Ising and $c=\frac{3}{2}$ BS. These
last ones are given in terms of the BS $|\alpha >$ for the charged
boson and the Ising ones $|f>$, $|\uparrow >$, $|\downarrow >$
according to (see ref.\cite{cft} for details):
\begin{align}
|\chi _{(0)}^{c=3/2}& >=|0>\otimes |\uparrow >+|2>\otimes |\downarrow > \\
|\chi _{(1)}^{c=3/2}& >=\frac{1}{2^{1/4}}\left( |1>+|3>\right) \otimes |f> \\
|\chi _{(2)}^{c=3/2}& >=|0>\otimes |\downarrow >+|2>\otimes |\uparrow >.
\end{align}
Such a factorization naturally arises already for the TM characters \cite%
{cgm2}.

The vacuum state for the TM model corresponds to the $\tilde{\chi}_{(0)}$
character which is the product of the vacuum state for the $c=\frac{3}{2}$
subtheory and that of the Ising one. From eqs. (\ref{vac1},\ref{vac3}) we
can see that the lowest energy state appears in two characters, so a linear
combination of them must be taken in order to define a unique vacuum state.
The correct BS in the untwisted sector are:
\begin{align}
|\tilde{\chi}_{((0,0),0)}& >=\frac{1}{\sqrt{2}}\left( |\tilde{\chi}%
_{(0)}^{+}>+|\tilde{\chi}_{(0)}^{-}>\right) =\sqrt{2}(|0>\otimes |\uparrow
\bar{\uparrow}>+|2>\otimes |\downarrow \bar{\uparrow}>)  \label{boud1} \\
|\tilde{\chi}_{((0,0),1)}& >=\frac{1}{\sqrt{2}}\left( |\tilde{\chi}%
_{(0)}^{+}>-|\tilde{\chi}_{(0)}^{-}>\right) =\sqrt{2}(|0>\otimes |\downarrow
\bar{\downarrow}>+|2>\otimes |\uparrow \bar{\downarrow}>) \\
|\tilde{\chi}_{((1,0),0)}& >=\frac{1}{\sqrt{2}}\left( |\tilde{\chi}%
_{(1)}^{+}>+|\tilde{\chi}_{(1)}^{-}>\right) =\sqrt{2}(|0>\otimes |\downarrow
\bar{\uparrow}>+|2>\otimes |\uparrow \bar{\uparrow}>) \\
|\tilde{\chi}_{((1,0),1)}& >=\frac{1}{\sqrt{2}}\left( |\tilde{\chi}%
_{(1)}^{+}>-|\tilde{\chi}_{(1)}^{-}>\right) =\sqrt{2}(|0>\otimes |\uparrow
\bar{\downarrow}>+|2>\otimes |\downarrow \bar{\downarrow}>) \\
|\tilde{\chi}_{(0)}(\varphi _{0})& >=\frac{1}{2^{1/4}}\left( |1>+|3>\right)
\otimes |D_{O}(\varphi _{0})>  \label{continous}
\end{align}
where we also added the states $|\tilde{\chi}_{(0)}(\varphi _{0})>$ in which
$|D_{O}(\varphi _{0})>$ is the continuous orbifold Dirichlet boundary state
defined in ref. \cite{Affleck}. For the special $\varphi _{0}=\pi /2$ value
one obtains:
\begin{equation}
|\tilde{\chi}_{(0)}>=\frac{1}{2^{1/4}}\left( |1>+|3>\right) \otimes |ff>.
\label{utgs}
\end{equation}
For the twisted sector we have:
\begin{align}
|\chi _{(0)}>& =\left( |0>+|2>\right) \otimes (|\uparrow \bar{f}%
>+|\downarrow \bar{f}>) \\
|\chi _{(1)}>& =\frac{1}{2^{1/4}}\left( |1>+|3>\right) \otimes (|f\bar{%
\uparrow}>+|f\bar{\downarrow}>).
\end{align}

Now, by using as reference state $|A>$ the vacuum state given in eq. (\ref%
{boud1}), we compute the chiral partition functions $Z_{AB}$ where $|B>$ are
all the BS just obtained \cite{noi1}:
\begin{eqnarray}
Z_{<\tilde{\chi}_{((0,0),0)}||\tilde{\chi}_{((0,0),0)}>} &=&\tilde{\chi}%
_{((0,0),0)}  \label{p1} \\
Z_{<\tilde{\chi}_{((0,0),0)}||\tilde{\chi}_{((1,0),0)}>} &=&\tilde{\chi}%
_{((1,0),0)}  \label{p2} \\
Z_{<\tilde{\chi}_{((0,0),0)}||\tilde{\chi}_{((0,0),1)}>} &=&\tilde{\chi}%
_{((0,0),1)}  \label{p3} \\
Z_{<\tilde{\chi}_{((0,0),0)}||\tilde{\chi}_{((1,0),1)}>} &=&\tilde{\chi}%
_{((1,0),1)}  \label{p4} \\
Z_{<\tilde{\chi}_{((0,0),0)}||\tilde{\chi}_{(0)}>} &=&\tilde{\chi}_{(0)}
\label{p5} \\
Z_{<\tilde{\chi}_{((0,0),0)}||\chi _{(0)}>} &=&\chi _{(0)}  \label{p6} \\
Z_{<\tilde{\chi}_{((0,0),0)}||\chi _{(1)}>} &=&\chi _{(1)}.  \label{p7}
\end{eqnarray}

So we can discuss topological order referring to the characters with the
implicit relation to the different boundary states present in the system.
Furthermore these BS can be associated to different kinds of linear defects,
which are compatible with conformal invariance \cite{noi1}.

\end{document}